\documentclass[nonacm, sigconf]{acmart}
\pdfoutput=1 

\usepackage{siunitx}
\usepackage{amsmath}
\usepackage{booktabs}   
\usepackage{subcaption}
\usepackage{graphicx}
\usepackage{multirow}
\usepackage[acronym]{glossaries}
\usepackage[abbreviations]{foreign}

\graphicspath{ {./img/} }

\newif\ifSuppressMemo
\ifSuppressMemo
\newcommand{\memo}[1]{}
\else
\usepackage{color}
\newcommand{\memo}[1]{{\bf \textcolor{red}{[#1]}}}
\fi

\AtBeginDocument{%
  \providecommand\BibTeX{{%
    \normalfont B\kern-0.5em{\scshape i\kern-0.25em b}\kern-0.8em\TeX}}}

\copyrightyear{2020}
\acmYear{2020}
\setcopyright{rightsretained}
\acmConference[NOSSDAV'20]{30th Workshop on Network and Operating System Support for Digital Audio and Video}{June 10--11, 2020}{Istanbul, Turkey}
\acmBooktitle{30th Workshop on Network and Operating System Support for Digital Audio and Video (NOSSDAV'20), June 10--11, 2020, Istanbul, Turkey}
\acmDOI{10.1145/3386290.3396933} 
\acmISBN{978-1-4503-7945-8/20/06}

\newacronym{m2p}{M2P}{motion-to-photon}
\newacronym{3dof}{3DoF}{three degrees of freedom}
\newacronym{6dof}{6DoF}{six degrees of freedom}
\newacronym{ran}{RAN}{Radio Access Network}
\newacronym{vr}{VR}{virtual reality}
\newacronym{ar}{AR}{augmented reality}
\newacronym{mr}{MR}{mixed reality}
\newacronym{xr}{XR}{extended reality}
\newacronym{hmd}{HMD}{head-mounted display}
\newacronym{mec}{MEC}{Mobile Edge Computing}
\newacronym{lod}{LOD}{level-of-detail}
\newacronym{fov}{FoV}{field-of-view}
\newacronym{qoe}{QoE}{Quality of Experience}
\newacronym{eeg}{EEG}{electroencephalogram}
\newacronym{emg}{EMG}{electromyography}
\newacronym{vp}{VP}{viewport}
\newacronym{sdp}{SDP}{Session Description Protocol}
\newacronym{ice}{ICE}{Interactive Connectivity Establishment}
\newacronym{ws}{WS}{WebSocket}
\newacronym{p2p}{P2P}{peer-to-peer}
\newacronym{slam}{SLAM}{Simultaneous Localization and Mapping}
\newacronym{slerp}{SLERP}{Spherical Linear Interpolation of Rotations}
\newacronym{rtt}{RTT}{round-trip time}
\newacronym{imu}{IMU}{inertial measurement unit}
\newacronym{mae}{MAE}{mean absolute error}
\newacronym{fps}{FPS}{frames per second}
\newacronym{rd}{RD}{rate-distortion}
\newacronym{autoreg}{AutoReg}{autoregressive}
\newacronym{aic}{AIC}{Akaike Information Criterion}
\newacronym{lat}{LAT}{look-ahead time}
\newacronym{hw}{\emph{hw}}{history window}

\begin{document}

\title{Low-latency Cloud-based Volumetric Video Streaming Using Head Motion Prediction}

\author{Serhan G{\"u}l}
\affiliation{%
  \institution{Fraunhofer HHI}
  \city{Berlin}
  \country{Germany}}
\email{serhan.guel@hhi.fraunhofer.de}

\author{Dimitri Podborski}
\affiliation{%
  \institution{Fraunhofer HHI}
  \city{Berlin}
  \country{Germany}}
\email{dimitri.podborski@hhi.fraunhofer.de}

\author{Thomas Buchholz}
\affiliation{%
  \institution{Deutsche Telekom AG}
  \city{Berlin}
  \country{Germany}}
\email{thomas.buchholz@telekom.de}

\author{Thomas Schierl}
\affiliation{%
  \institution{Fraunhofer HHI}
  \city{Berlin}
  \country{Germany}}
\email{thomas.schierl@hhi.fraunhofer.de}

\author{Cornelius Hellge}
\affiliation{%
  \institution{Fraunhofer HHI}
  \city{Berlin}
  \country{Germany}}
\email{cornelius.hellge@hhi.fraunhofer.de}

\renewcommand{\shortauthors}{G{\"u}l and Podborski, et al.}


\begin{abstract}
Volumetric video is an emerging key technology for immersive representation of 3D spaces and objects. Rendering volumetric video requires lots of computational power which is challenging especially for mobile devices. To mitigate this, we developed a streaming system that renders a 2D view from the volumetric video at a cloud server and streams a 2D video stream to the client. However, such network-based processing increases the motion-to-photon (M2P) latency due to the additional network and processing delays. In order to compensate the added latency, prediction of the future user pose is necessary. We developed a head motion prediction model and investigated its potential to reduce the M2P latency for different look-ahead times. Our results show that the presented model reduces the rendering errors caused by the M2P latency compared to a baseline system in which no prediction is performed.
\end{abstract}


\maketitle

\section{Introduction}
Recent advances in hardware for displaying of immersive media have aroused a huge market interest in \gls{vr} and \gls{ar}  applications. Although the initial interest was focused on omnidirectional (\ang{360}) video applications, with the improvements in capture and processing technologies, volumetric video has recently started to become the center of attention~\cite{schreer2019}. Volumetric videos capture the 3D space and objects and enable services with \gls{6dof}, allowing a viewer to freely change both the position in space and the orientation.

Although the computing power of mobile end devices has dramatically increased in the recent years, rendering rich volumetric objects is still a very demanding task for such devices. Moreover, there are yet no efficient hardware decoders for volumetric content (e.g. point clouds or meshes), and software decoding can be prohibitively expensive in terms of battery usage and real-time rendering requirements. 
One way of decreasing the processing load on the client is to avoid sending the volumetric content and instead send a 2D rendered view corresponding to the position and orientation of the user. To achieve this, the expensive rendering process needs to be offloaded to a server infrastructure.
Rendering 3D graphics on a powerful device and displaying the results on a \emph{thin} client connected through a network is known as remote (or interactive) rendering~\cite{shi2015}. Such rendering servers can be deployed at a cloud computing platform such that the resources can be flexibly allocated and scaled up when more processing load is present.

In a cloud-based rendering system, the server renders the 3D graphics based on the user input (e.g, head pose) and encodes the rendering result into a 2D video stream. Depending on the user interaction, camera pose of the rendered video is dynamically adapted by the cloud server. After a matching view has been rendered and encoded, the obtained video stream is transmitted to the client. The client can efficiently decode the video using its hardware video decoders and display the video stream. Moreover, network bandwidth requirements are reduced by avoiding the transmission of the volumetric content.

Despite these advantages, one major drawback of cloud-based rendering is an increase in the end-to-end latency of the system, typically known as \gls{m2p} latency. Due to the added network latency and processing delays (rendering and encoding), the amount of time until an updated image is presented to the user is greater than a local rendering system. It is well-known that an increase in \gls{m2p} latency may cause an unpleasant user experience and motion sickness~\cite{adelstein2003, allison2001}. 
One way to reduce the network latency is to move the volumetric content to an \emph{edge} server geographically closer to the user. Deployment of real-time communication protocols such as WebRTC are also necessary for ultra-low latency video streaming applications~\cite{holmberg2015}. The processing latency at the rendering server is another significant latency component. Therefore, using fast hardware-based video encoders is critical for reducing the encoding latency.

Another way of reducing the \gls{m2p} latency is to predict the future user pose at the remote server and send the corresponding rendered view to the client. Thus, it is possible to reduce or even completely eliminate the \gls{m2p} latency, if the user pose is predicted for a \gls{lat} equal to or larger than the M2P latency of the system~\cite{bao2016}. However, mispredictions of head motion may potentially degrade the user's \gls{qoe}. Thus, design of accurate prediction algorithms has been a popular research area, especially for the viewport prediction for \ang{360} videos (see Section~\ref{sec:bg_prediction}). However, application of such algorithms to \gls{6dof} movement (i.e. translational and rotational) has not yet been investigated. 

In this paper, we describe our cloud-based volumetric streaming system, present a prediction model to forecast the \gls{6dof} position of the user and investigate the achieved rendering accuracy using the developed prediction model. Additionally, we present an analysis of the latency contributors in our system and a simple latency measurement technique that we used to characterize the the \gls{m2p} latency of our system.
\section{Background}
\label{sec:background}
\begin{figure*}[tb]
    \centering
    \includegraphics[width=0.9\linewidth]{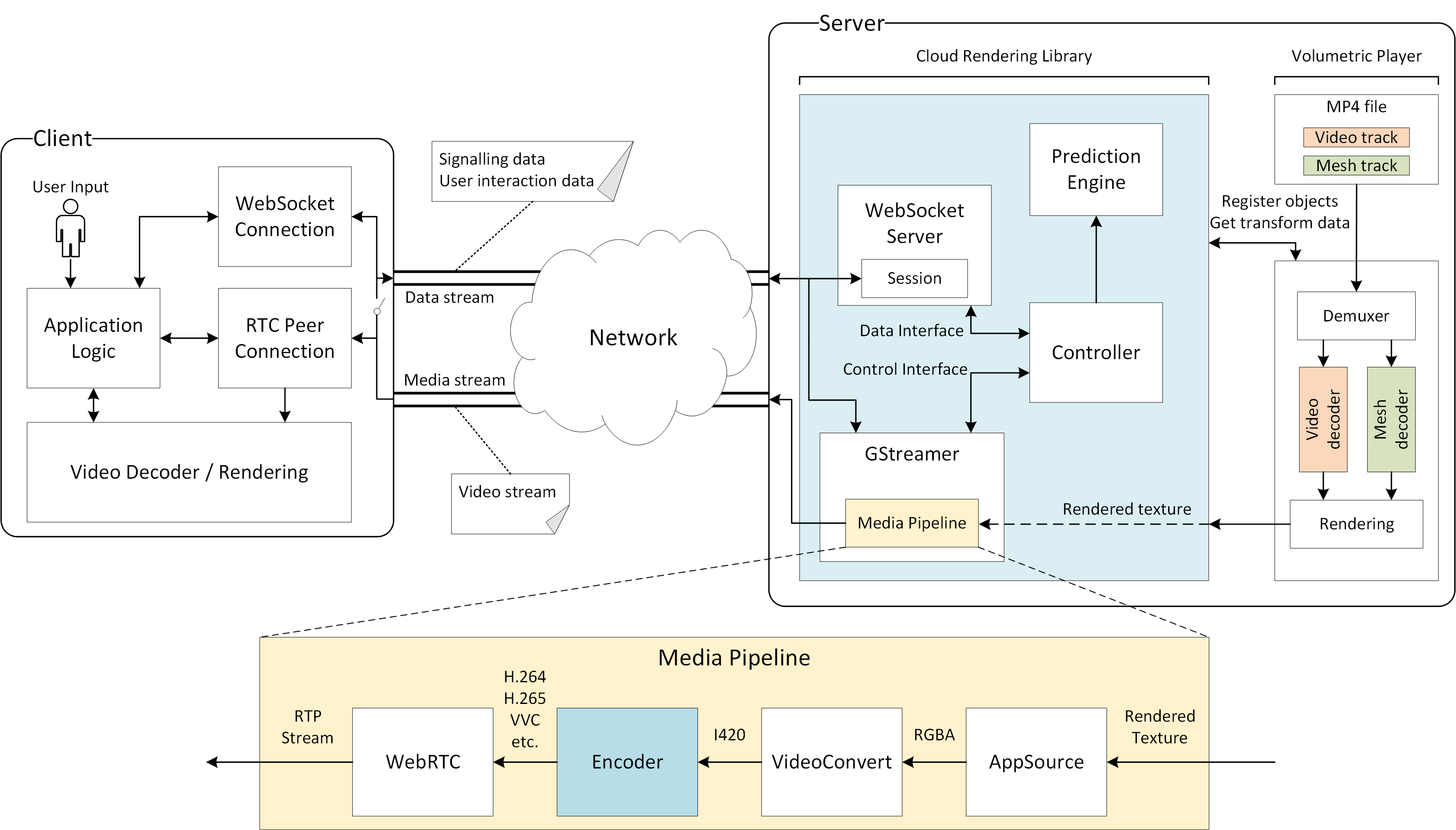}
    \caption{Overview of the system components and interfaces.}
    \label{fig:overview}
\end{figure*}
\subsection{Volumetric video streaming}
Some recent works present initial frameworks for streaming of volumetric videos. 
Qian et al.~\cite{qian2019} developed a proof-of-concept point cloud streaming system and introduced optimizations to reduce the \gls{m2p} latency.
Van der Hooft et al.~\cite{van2019} proposed an adaptive streaming framework compliant to the recent point cloud compression standard MPEG V-PCC~\cite{schwarz2018}. They used their framework for HTTP adaptive streaming of scenes with multiple dynamic point cloud objects and presented a rate adaptation algorithm that considers the user's position and focus.
Petrangeli et al.~\cite{petrangeli2019} proposed a streaming framework for \gls{ar} applications that dynamically decides which virtual objects should be fetched from the server as well as their \glspl{lod}, depending on the proximity of the user and likelihood of the user to view the object. 

\subsection{Cloud rendering systems}
The concept of remote rendering was first put forward to facilitate the processing of 3D graphics rendering when PCs did not have sufficient computational power for intensive graphics tasks. A detailed survey of interactive remote rendering systems in the literature is presented in ~\cite{shi2015}. 
Shi et al.~\cite{shi2019} proposed a \gls{mec} system to stream \gls{ar} scenes containing only the user's \gls{fov} and a latency-adaptive margin around the \gls{fov}. They evaluate the performance of their prototype on a \gls{mec} node connected to a 4G (LTE) testbed.
Mangiante et al.~\cite{mangiante2017} proposed an edge computing framework that performs \gls{fov} rendering of \ang{360} videos. Their system aims to optimize the required bandwidth as well as reduce the processing requirements and battery utilization.

Cloud rendering has also started to receive increasing interest from the industry, especially for cloud gaming services. Nvidia CloudXR~\cite{nvidia2019} provides an SDK to run computationally intensive \gls{xr} applications on Nvidia cloud servers to deliver advanced graphics performances to thin clients. 

\subsection{Head motion prediction techniques}
\label{sec:bg_prediction}
Several sensor-based methods have been proposed in the literature that attempt to predict the user's future viewport for optimized streaming of \ang{360}-videos.
Those can be divided into two categories. Works such as~\cite{bao2016, bao2017, sanchez2019, aykut2018, oculus16} were specifically designed for \gls{vr} applications and use the sensor data from \glspl{hmd} whereas the works in~\cite{barniv2005, koirala2015, bai2011} attempt to infer user motion based on some physiological data such as \gls{eeg} and \gls{emg} signals.
Bao et al.~\cite{bao2016} collected head orientation data and exploited the correlations in different dimensions to predict the head motion using regression techniques. Their findings indicate that \glspl{lat} of \SIrange{100}{500}{ms} is a feasible range for sufficient prediction accuracy.
Sanchez et al.~\cite{sanchez2019} analyzed the effect of \gls{m2p} latency on a tile-based streaming system and proposed an angular acceleration-based prediction method to mitigate the impact on the observed fidelity. 
Barniv et al.~\cite{barniv2005} used the myoelectric signals obtained from \gls{emg} devices to predict the impending head motion. They trained a neural network to map \gls{emg} signals to trajectory outputs and experimented with combining \gls{emg} output with inertial data. Their findings indicate that a \glspl{lat} of \SIrange{30}{70}{ms} are achievable with low error rates. 

Most of the previous works target \gls{3dof} \gls{vr} applications and thus focus on prediction of only the head orientation in order to optimize the streaming \ang{360} videos. However, little work has been done so far on prediction of \gls{6dof} movement for advanced \gls{ar} and \gls{vr} applications. In Sec.~\ref{sec:prediction}, we present an initial statistical model for \gls{6dof} prediction and discuss our findings.

\section{System Architecture}
\label{sec:system}
This section presents the system architecture of our cloud rendering-based volumetric video streaming system and describes its different components. A simplified version of this architecture is shown in Fig.~\ref{fig:overview}. 

\subsection{Server architecture}
\label{sec:server_arch}
The server-side implementation is composed of two main parts: a volumetric video player and a cross-platform cloud rendering library, each described further in more detail.
\subsubsection*{Volumetric video player} 

The volumetric video player is implemented in Unity and plays a single MP4 file which has one video track containing the compressed texture data and one mesh track containing the compressed mesh data of a volumetric object.
Before the playback of a volumetric video starts, the player registers all the required objects. For example, the virtual camera of the rendered view and the volumetric object are registered, and those can later be controlled by the client. After initialization, the volumetric video player can start playing the MP4 file. During playout, both tracks are demultiplexed and fed into the corresponding decoders; video decoder for texture track and mesh decoder for mesh track. After decoding, each mesh is synchronized with the corresponding texture and rendered to a scene.
The rendered view of the scene is represented by a Unity \emph{RenderTexture} that is passed to our cloud rendering library for further processing.
While rendering the scene, the player concurrently asks the cloud rendering library for the latest positions of the relevant objects that were previously registered in the initialization phase.

\subsubsection*{Cloud rendering library} 
We created a cross-platform cloud rendering library written in C++ that can be integrated into a variety of applications. In the case of the Unity application, the library is integrated into the player as a native plugin.
The library utilizes the GStreamer WebRTC plugin for low-latency video streaming between the server and client that is integrated into a media pipeline as described in Sec.~\ref{sec:media_pipeline}. In addition, the library provides interfaces for registering the objects of the rendered scene and retrieving the latest client-controlled transformations of those objects while rendering the scene. In the following, we describe the modules of our library, each of which runs asynchronously in its own thread to achieve high performance.

The \textbf{WebSocket Server} is used for exchanging signaling data between the client and the server. Such signaling data includes \gls{sdp}, \gls{ice} as well as application-specific metadata for scene description. In addition, \gls{ws} connection can also be used for sending the control data, e.g. changing the position and orientation of any registered game object or camera.
Both, plain WebSockets as well as Secure WebSockets are supported which is important for practical operation of the system.

The \textbf{GStreamer} module contains the media processing pipeline which takes the rendered texture and compressed it into a video stream that is sent to the client using the WebRTC plugin. The most important components of the media pipeline are described in Sec.~\ref{sec:media_pipeline}.

The \textbf{Controller} module represents the application logic and controls the other modules depending on the application state. For example, it closes the media pipeline if the client disconnects, re-initializes the media pipeline when a new client has connected, and updates the controllable objects based on the output of the Prediction Engine.

The \textbf{Prediction Engine} implements a regression-based prediction method (please refer to Sec.~\ref{sec:prediction}) and provides interfaces for usage of other potential methods. Based on the previously received input from the client and the implemented algorithm, the module updates the position of the registered objects accordingly such that the rendered scene corresponds to the predicted positions of the object after a given \gls{lat}.

\subsection{Client architecture}
\label{sec:client_arch}
The client-side architecture is depicted on the left side of the Fig.~\ref{fig:overview}. Before the streaming session starts, the client establishes a \gls{ws} connection to the server and asks the server to send a description of the rendered scene. The server responds with a list of objects and parameters which the client is later allowed to update. After receiving the scene description, the client replicates the scene and initiates a \gls{p2p} WebRTC connection to the server.
The server and client begin the WebRTC negotiation process by sending \gls{sdp} and \gls{ice} data over the established \gls{ws} connection. Finally, the \gls{p2p} connection is established, and the client starts receiving a video stream corresponding to the current view of the volumetric video. At the same time, the client can use the \gls{ws} connection, as well as the RTCPeerConnection for sending control data to the server in order to modify the properties of the scene. For example, the client may change its \gls{6dof} position, or it may rotate, move and scale any volumetric object in the scene.

We have implemented both a web player in JavaScript and a native application for the HoloLens, the untethered \gls{ar} headset from Microsoft. While our web application targets \gls{vr}, our HoloLens application is implemented for \gls{ar} use cases.
In the HoloLens application, we perform further processing to remove the background of the video texture before rendering the texture onto the \gls{ar} display. In general, the client-side architecture remains the same for both \gls{vr} and \gls{ar} use cases, and the most complex client-side module is the video decoder. Thus, the complexity of our system is concentrated largely in our cloud-based rendering server.

\subsection{Media pipeline}
\label{sec:media_pipeline}
The simplified structure of the media pipeline is shown in the bottom part of Fig.~\ref{fig:overview}. A rendered texture is given to the media pipeline as input using the \emph{AppSource} element of Gstreamer. Since the rendered texture is originally in RGB format but the video encoder requires YUV input, we use the \emph{VideoConvert} element to convert the RGB texture to I420 format\footnote{https://www.fourcc.org/pixel-format/yuv-i420}. After conversion, the texture is passed to the encoder element, which can be set to any supported encoder on the system. 
Since encoder latency is a significant contributor the overall \gls{m2p} latency, we evaluated the encoding performances of different encoders for a careful selection. For detailed results on the encoder performance, please refer to Sec.~\ref{sec:enc_latency}.

After the texture is encoded the resulting video bitstream is packaged into RTP packets, encrypted and sent to the client using WebRTC. WebRTC was chosen as the delivery method since it allows us to achieve an ultra-low latency while using the \gls{p2p} connection between the client and server. In addition, WebRTC is already widely adopted by different web browsers allowing our system to support several different platforms.
\section{Motion-to-Photon Latency}
\label{sec:latency}
\begin{figure}[htbp]
  \centering
  \includegraphics[width=0.9\linewidth]{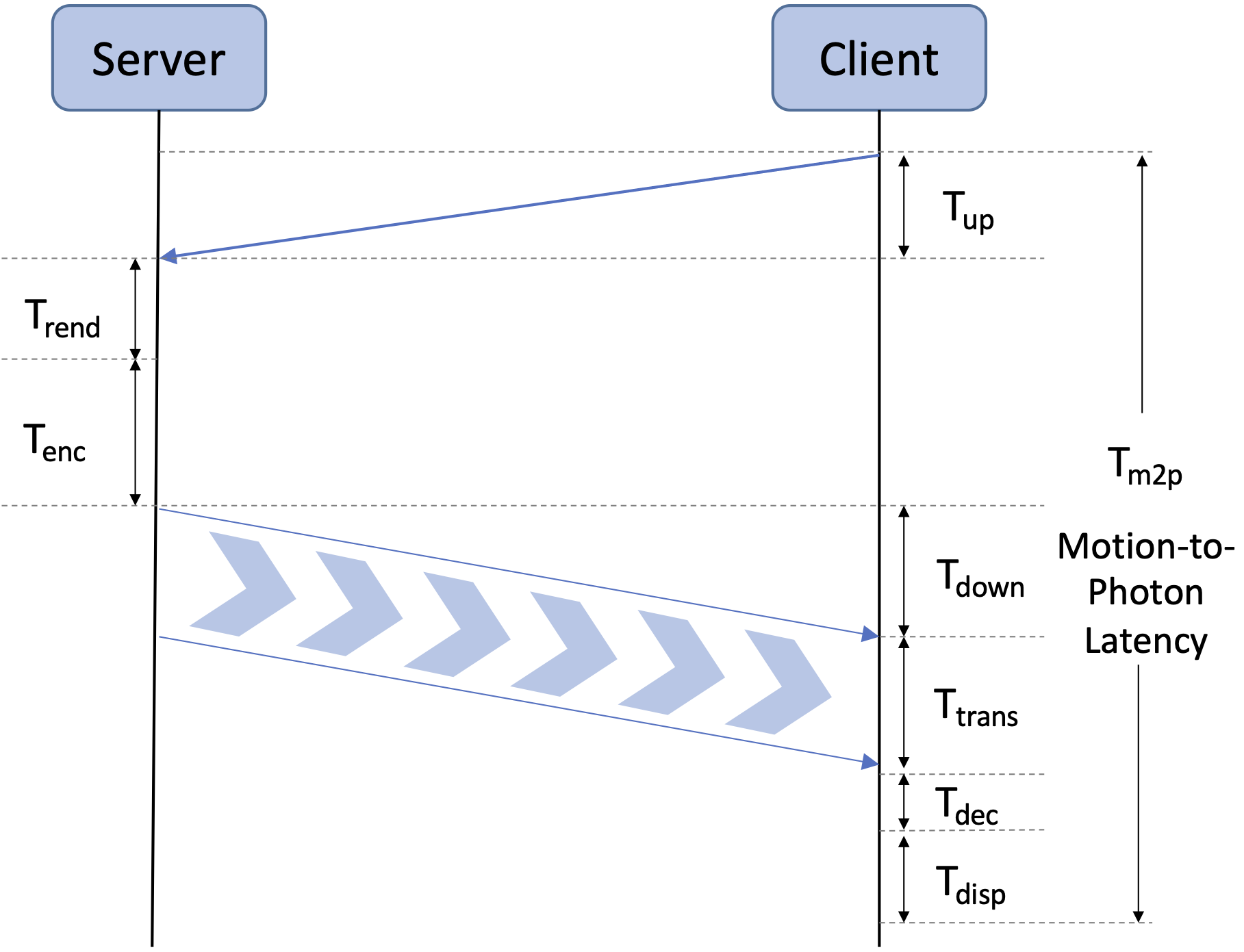}
  \caption{Components of the \acrlong{m2p} latency for a remote rendering system.}
  \label{fig:latency}
\end{figure}

The different components of the \gls{m2p} latency are illustrated in Fig.~\ref{fig:latency} and related by
\begin{equation}
    \mathrm{T_{M2P}} = \mathrm{T_{server}} + \mathrm{T_{network}} + \mathrm{T_{client}}
    \label{eq:e2e}
\end{equation}
where $T_{server}$, $T_{client}$ and $T_{network}$ consist of the following component latencies:
\begin{gather}
\mathrm{T_{server}} = \mathrm{T_{rend}} + \mathrm{T_{enc}} \\
\mathrm{T_{network}} = \mathrm{T_{up}} + \mathrm{T_{down}} + \mathrm{T_{trans}} \\
\mathrm{T_{client}} = \mathrm{T_{dec}} + \mathrm{T_{disp}}
\end{gather}

In our analysis, we neglect the time for the \gls{hmd} to compute the user pose using its tracker module. This computation is typically based on a fusion of the sensor data from the \glspl{imu} and visual data from the cameras. Although \gls{ar} device cameras typically operate at \SIrange{30}{60}{Hz}, \glspl{imu} are much faster and a reliable estimation of the user pose can be performed with a frequency of multiple \si{\kilo\Hz}~\cite{lincoln2016, daqri2018}. Thus, the expected tracker latency is on the order of microseconds. 

$\mathrm{T_{enc}}$ is the time to compress a frame and depends on the encoder type (hardware or software) and the picture resolution. We present a detailed latency analysis of different encoders that we tested for our system in Section~\ref{sec:enc_latency}.

$\mathrm{T_{network}}$ is the network \gls{rtt}. It consists of the propagation delay ($\mathrm{T_{up}}+\mathrm{T_{down}}$) and the transmission delay $\mathrm{T_{trans}}$. $\mathrm{T_{up}}$, the time for the server to retrieve sensor data from the client, and $\mathrm{T_{down}}$ the time for the server to transmit a compressed frame to the client.

$\mathrm{T_{rend}}$ is the time for the server to generate a new frame by rendering a view from the volumetric data based on the actual user pose.
In general, it can be set to match the frame rate of the encoder for a given rendered texture resolution.

$\mathrm{T_{dec}}$ is the time to decode a compressed frame on the client device and is typically much smaller than $\mathrm{T_{enc}}$ since video decoding is inherently a faster operation than video encoding. Also, the end devices typically have hardware-accelerated video decoders that further reduce the decoding latency.

$\mathrm{T_{disp}}$ is the display latency and mainly depends on the refresh rate of the display. For a typical refresh rate of \SI{60}{Hz}, the average value of $\mathrm{T_{disp}}$ is \SI{8.3}{ms}, and the worst-case value is \SI{16.6}{ms} in case the decoded frame misses the current VSync signal and has to wait in the frame buffer for the next VSync signal.

\subsection{Encoder latency}
\label{sec:enc_latency}
We characterized the encoding speeds of different encoders using the test dataset provided by ISO/ITU Joint Video Exploration Team (JVET) for the next generation video coding standard Versatile Video Coding (VVC)~\cite{jvet_VVC_CfP}. 
In our measurements, we used the FFmpeg libraries of the encoders NVENC, x264, x265, and Intel SVT-HEVC, enabled their low-latency presets and measured the encoded \gls{fps} using FFmpeg \emph{-benchmark} option. The measurements were performed on a Ubuntu 18.04 machine with 16 Intel Xeon Gold 6130 CPU (2.10GHz) CPUs using the default threading options for the tested software-based encoders.
For x264 and x265, we used the \emph{ultrafast} preset and \emph{zerolatency} tuning \cite{ffmpeg_x264}. For NVENC, we evaluated the presets default, high-performance (HP), low-latency (LL) and low-latency high-performance (LLHP). A brief description of the NVENC presets can be found in~\cite{twitch2018}. 

Table~\ref{table:encoder} shows the mean \gls{fps} over all tested sequences for different encoders. We observed that both H.264 and HEVC encoders of NVENC are significantly faster than x264 andx265 (both using ultrafast preset and zerolatency tuning) as well as SVT-HEVC (Low delay P). NVENC is able to encode 1080p and 4K videos in our test dataset with encoding speeds up to \SI{800}{fps} and \SI{200}{fps}, respectively. We also observed that for some sequences, HEVC encoding turned out to be faster than H.264 encoding. We believe that this difference is caused by a more efficient GPU implementation for HEVC.

All the low-latency presets tested in our experiments turn off B-frames to reduce latency.
Despite that, we observed that the picture quality obtained by NVENC in terms of PSNR is comparable to the other tested encoders (using low-latency presets). As a result of our analysis, we decided to use NVENC H.264 (HP preset) in our system.

\begin{table}[htbp]
\setlength{\tabcolsep}{4pt} 
\renewcommand{\arraystretch}{0.8} 
\small
\caption{Mean encoding performances over all tested sequences for different encoders and presets.}
\begin{tabular}{@{}llll@{}}
\toprule
Standard & Encoder & Preset & Mean \gls{fps}  \\ 
\midrule
\multirow{5}{*}{H.264} 
& x264 & Ultrafast & 81  \\
& NVENC  & Default  & 353 \\
& NVENC  & HP & \textbf{465} \\
& NVENC  & LL  & 359 \\
& NVENC  & LLHP  & 281  \\
\midrule
\multirow{5}{*}{HEVC} 
& x265   & Ultrafast  & 33 \\
& SVT-HEVC  & Low delay P & 74  \\
& NVENC  & Default & 212 \\
& NVENC  & HP  & \textbf{492} \\
& NVENC  & LL  & 278  \\
& NVENC  & LLHP   & 211  \\
\bottomrule
\end{tabular}
\label{table:encoder}
\vspace{-4mm}
\end{table}

\subsection{Latency measurements}
\label{sec:m2p}
We developed a framework to measure the \gls{m2p} latency of our system. In our setup, we run the server application on an Amazon EC2 instance in Frankfurt, and the client application runs in a web browser in Berlin which is connected to the Internet over WiFi.

We implemented a server-side console application which is using the same cloud rendering library as described in Sec.~\ref{sec:server_arch} but instead of sending the rendered textures from the volumetric video player, the application sends predefined textures (known by the client) depending on the received control data from the client. These textures consist of simple vertical bars with different colors. For example, if the client instructs the server application to move the main camera to position $P_1$, the server pushes the texture $F_1$ into the media pipeline. Similarly, another camera position $P_2$ results in the texture $F_2$. 

On the client side, we implemented a web-based application that connects to the server application and renders the received video stream to a canvas. Since the client knows exactly how those textures look like, it can evaluate the incoming video stream and determine when the requested texture was rendered on the screen. As soon as the client application sends $P_1$ to the server, it starts the timer and checks the canvas for $F_1$ at every web browser window \emph{repaint} event.
According to the W3C recommendation~\cite{w3c2015}, the repaint event matches the refresh rate of the display.
As soon as the texture $F_1$ is detected the client stops the timer and computes the \gls{m2p} latency $\mathrm{T_{M2P}}$.

Once the connection is established, the user can start the session by defining the number of independent measurements. Since we are using the second smallest instance type of Amazon EC2 (t2.micro), we set the size of each video frame to $512\times512$ pixels. We encode the stream using x264 configured with ultrafast preset and zerolatency tuning with an encoding speed of \SI{\sim 80}{fps}. As an example, we set the client to perform 100 latency measurements and calculated the average, minimum and maximum \gls{m2p} latency. 
Our results show that $\mathrm{T_{M2P}}$ fluctuates between \SI{41}{ms} and \SI{63}{ms}, and the measured average \gls{m2p} latency is \SI{58}{ms}.

\setcounter{footnote}{0}
\section{Head motion prediction}
\label{sec:prediction}

\begin{figure*}[htb]
\centering
\includegraphics[width=\linewidth]{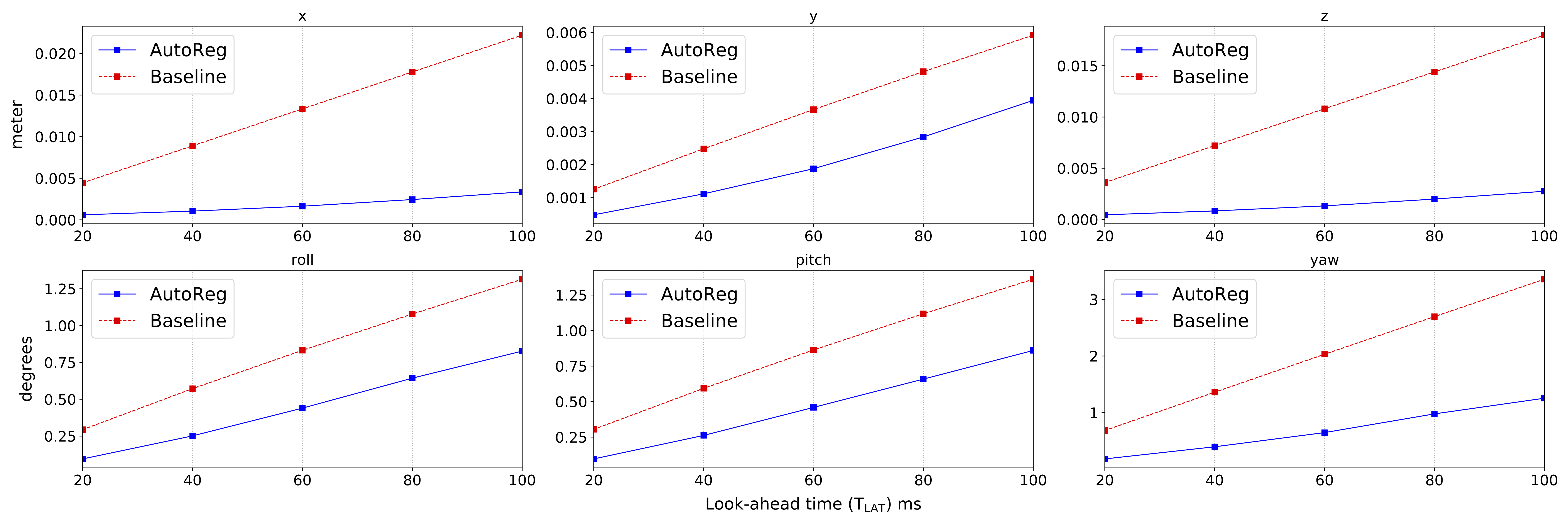}
\caption{Mean absolute error (MAE) for different translational and rotational components averaged over five users. Results are given for the look-ahead times $\mathrm{T_{LAT}}$ in the range \SIrange{20}{100}{ms}.}
\label{fig:mean_results}
\end{figure*}

One important technique to mitigate the increased \gls{m2p} latency in a cloud-based rendering system is the prediction of the user's future pose. In this section, we describe our statistical prediction model for \gls{6dof} head motion prediction and evaluate its performance using real user traces.

\subsection{Data collection}
\label{sec:data_collection}
We collected motion traces from five users while they were freely interacting with a static virtual object using Microsoft HoloLens. We recorded the users' movements in 6DoF space; i.e., collected position samples ($x$, $y$, $z$) and rotation samples represented as quaternions ($q_x$, $q_y$, $q_z$, $q_w$). Since the raw sensor data we obtained from HoloLens was unevenly sampled (i.e. different temporal distances between consecutive samples) at \SI{60}{Hz}, we interpolated the data to obtain temporally equidistant samples. We upsampled the position data using linear interpolation and the rotation data (quaternions) using \gls{slerp}~\cite{shoemake1985}. Thus, we obtained an evenly-sampled dataset with a sampling rate of \SI{200}{Hz} (one sample at each \SI{5}{ms}).\footnote{Our \gls{6dof} head movement dataset is freely available on Github for further usage in research community under: \url{https://github.com/serhan-gul/dataset_6DoF}} 

\subsection{Prediction method}
\label{sec:method}
We use a simple \gls{autoreg} model to predict the future user pose based on a time series of its past values. \gls{autoreg} models use a linear combination of the past values of a variable to forecast its future values~\cite{hyndman2018}.

An \gls{autoreg}  model of lag order $\rho$ can be written as
\begin{equation}
\label{eq:autoreg}
    y_t = c+\phi_1 y_{t-1}+\phi_2 y_{t-2}+\dots+\phi_\rho y_{\rho-1}+\epsilon_t
\end{equation}
where $y_t$ is the true value of the time series $y$ at time $t$, $\epsilon_t$ is the white noise, $\phi_i$ are the coefficients of the model. Such a model with $\rho$ lagged values is referred to as an $\mathrm{AR}(\rho)$ model. Some statistics libraries can determine the lag order automatically using statistical tests such as the \gls{aic}~\cite{akaike1973}.

We used the $x$ and $q_x$ values from one of the collected traces as training data and created two \gls{autoreg} models using the Python library \emph{statsmodels}~\cite{statsmodels}, for translational and rotational components, respectively. Our model has a lag order of 32 samples i.e. it considers a \gls{hw} of the past $32*5=160$~ms and predicts the next sample using \eqref{eq:autoreg}. Typically we need to predict not only the next sample but multiple samples in the future to achieve a given \gls{lat}; therefore, we repeat the prediction step by adding the just-predicted sample to the history window and iterating \eqref{eq:autoreg} until we obtain the future sample corresponding to the desired \gls{lat}. The process is then repeated for each frame e.g. each \SI{10}{ms} for an assumed \SI{100}{Hz} display refresh rate.

We used the trained model to predict the users' translational ($x$, $y$, $z$), and rotational motion ($q_x$, $q_y$, $q_z$, $q_w$). We perform the prediction of rotations in the quaternion domain 
since we readily obtain quaternions from the sensors and they allow smooth interpolation using techniques like \gls{slerp}. After prediction, we convert the predicted quaternions to Euler angles (yaw, pitch, roll) and evaluate the prediction accuracy in the domain of Euler angles since they are better suited for understanding the rendering offsets in terms of angular distances.

\subsection{Evaluation}
In our evaluation, we investigated the effect of prediction on the accuracy of the rendered image displayed to the user, i.e. the rendering offset. Specifically, we compared the predicted user pose (given a certain \acrlong{lat} $\mathrm{T_{LAT}}$) to the real user pose as obtained from the sensors. As a benchmark, we evaluated a baseline case in which the rendered pose lags behind the actual user pose by a delay corresponding to the \gls{m2p} latency ($\mathrm{T_{M2P}}$), i.e., no prediction is performed. 

For each user trace, we evaluated the prediction algorithm for a $\mathrm{T_{LAT}}$ ranging between \SIrange{20}{100}{ms}. In each experiment, we assume that the \gls{m2p} latency is equal to the prediction time ($\mathrm{T_{LAT}}=\mathrm{T_{M2P}}$) such that the prediction model attempts to predict the pose that the user will attain at the time the rendered image is displayed to the user. We evaluated our results by computing the \gls{mae} between the true and predicted values for the different components.

\begin{figure}[htb]
\centering
\includegraphics[width=\linewidth]{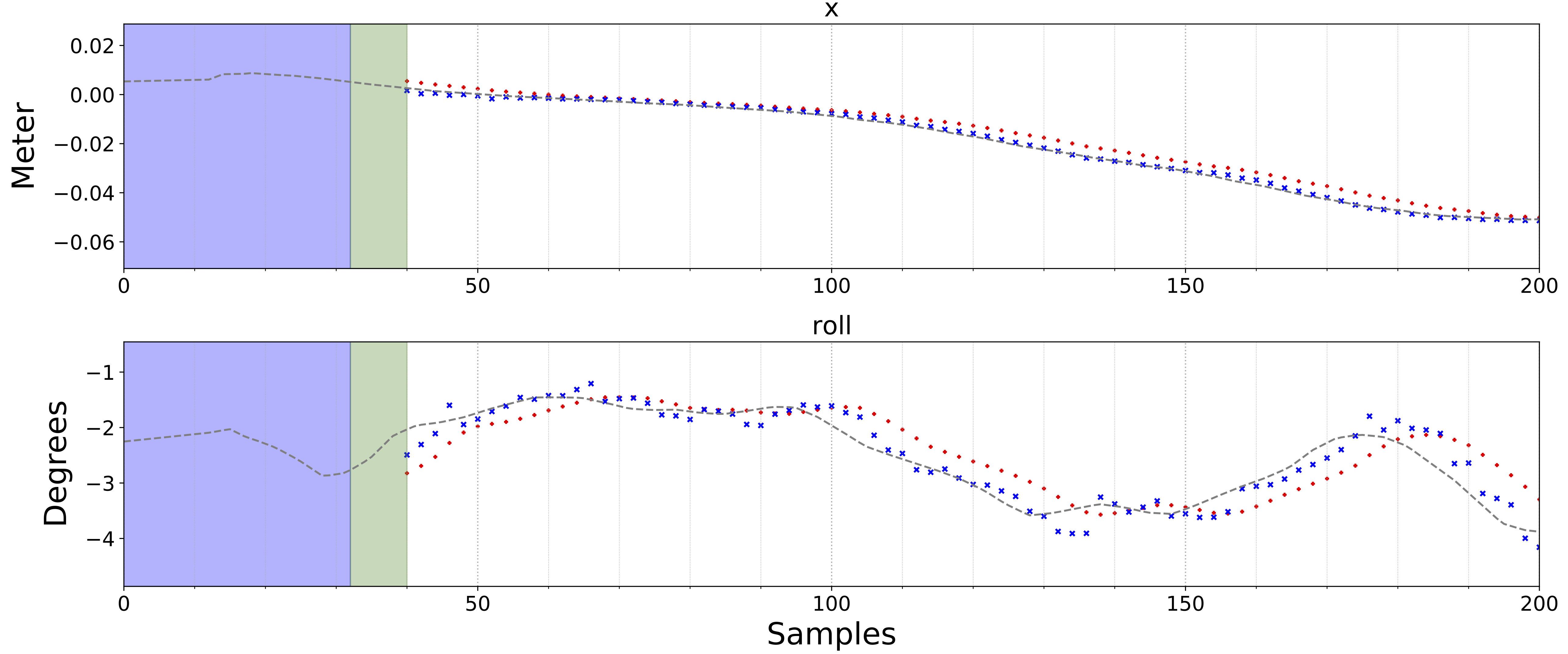}
\caption{Comparison of the prediction (blue) and baseline (red) results for the \emph{x} and \emph{roll} components of one of the traces (sample time \SI{5}{ms}; showing the time range \SIrange{0}{1}{s}) for $\mathrm{T_{LAT}}=40$~ms. The dashed gray line shows the recorded sensor data.}
\label{fig:single_trace}
\end{figure}

Fig.~\ref{fig:mean_results} compares the average rendering errors over five traces obtained using our prediction method to the baseline. We observe that for all considered $\mathrm{T_{LAT}}$, prediction reduces the average rendering error for both positional and rotational components.

Fig.~\ref{fig:single_trace} shows for one of the traces the predicted and baseline (lagged by \gls{m2p} latency) values for the \emph{x} and \emph{roll} components. At the beginning of the session, the prediction module collects the required amount of samples for a \gls{hw} of \SI{160}{ms} and makes the first prediction, i.e., the pose that the user is predicted to attain after a time of $\mathrm{T_{LAT}}=40$~ms (green shaded). We observe that the accuracy of the prediction depends on the frequency of the abrupt, short-term changes of the user pose. If a component of the user pose linearly changes over a \gls{hw} (without changing direction), the resulting predictions for that component are fairly accurate. Otherwise, if short-term changes are present within a \gls{hw}, the prediction tends to perform worse than the baseline.

\section{Conclusion}
We presented a cloud-based based volumetric streaming system that offloads the rendering to a powerful server and thus reduces the rendering load on the client-side. 
To compensate the added network and processing latency, we developed a method to predict the user's head motion in ~\acrlong{6dof}. Our results show that the developed prediction model reduces the rendering errors caused by the added latency due to the cloud-based rendering. 
In our future work, we will analyze the effect of motion-to-photon latency on the user experience through subjective tests and develop more advanced prediction techniques e.g. based on Kalman filtering.

\bibliographystyle{ACM-Reference-Format}
\bibliography{mmsys20-ref.bib}


\end{document}